\documentstyle[12pt]{article}
\begin{document} 

\title{Nonlinear Interaction of Transversal Modes in a $CO_2$ Laser}
\author{ Ricardo L\'{o}pez-Ruiz, Gabriel B. Mindlin\\
  Carlos  P\'{e}rez-Garc\'{\i}a\\
 	\\
\small Universidad Privada de Navarra\\ 
\small Departamento de F\'{\i}sica y Matem\'{a}tica Aplicada \\
\small E-31080 Pamplona, Navarra, Spain \\
\small and	   \\ \\
 Jorge R. Tredicce\\
	   \\
\small Institute Non Lineaire de Nice\\
\small F-06034, Nice Cedex, France}
\date{ }

\maketitle
\baselineskip 8mm

\begin{center} {\bf Abstract} \end{center}
We show the possibility of achieving experimentally a Takens-Bogdanov
bifurcation for the nonlinear interaction of two transverse modes ($l=\pm{1}$)
in a $CO_2$ laser. The system has a basic $O(2)$ symmetry which is 
perturbed by some symmetry-breaking effects that still preserve 
the $Z_2$ symmetry. The pattern dynamics near this codimension two bifurcation
 under such symmetries is described.
This dynamics changes drastically when the laser properties are modified.
\newline
PACS number(s):42.65.Pc, 02.20.-a, 42.55.Lt

\newpage 
\section{Introduction}

Pattern formation in physical systems is an area of active research. It has 
been recently pointed that lasers can provide a "test bench" for these
 studies, 
as it is possible to have many active transverse modes which, through nonlinear 
interaction, give a complicated spatiotemporal dynamics
 \cite{green}. Recent evidence shows that some interesting phenomena  
such as defects, vortices, chaotic alternancy, etc., can be found in laser
 systems \cite{arecchi}. Unfortunately, in most cases these works do not 
enlighten the mechanisms behind the transition from simple solutions to the
spatiotemporal uncorrelated ones. A first step in understanding this
complex dynamics  is to
study the nonlinear interaction among a few modes which bifurcate from the 
zero solution \cite{dangelo,lopez}. This article strives in this direction.
\par
It has been reported that some qualitative features of the transverse
 patterns of intensity observed in $CO_2$ lasers can be explained in the frame
of the theory 
of bifurcations in the presence of symmetries \cite{green,golubiski}.
The laser tube imposes the $O(2)$ symmetry (rotations and reflections).
The mode amplitude equations having this symmetry predict that the  stable
modes with non-zero  angular momentum  bifurcating from zero should be 
travelling waves, in contradiction with experiments.
Anisotropies in the laser parameters (pumping, losses or  disalignments in
 the setup,etc.) break  the $O(2)$ symmetry.
An agreement between theory and experiment can be reached by introducing a  
 symmetry-breaking term  preserving the $Z_2$ symmetry.
This model succesfully predicts the stability of the standing waves widely 
observed in the laser transverse section.
The prediction of secondary solutions that arise from the
 standing waves is another achievement of the model with 
"imperfect" symmetry. These solutions  are either a mixture of a standing wave and 
travelling waves, or modulated waves that appear from a Hopf bifurcation
of the standing waves  \par

	The conditions for a $CO_2$ laser displaying those solutions
that bifurcate simultaneously from the standing wave (codimension two point)
 are analyzed here. The dynamics near such
bifurcation and the intensity patterns resulting from this situation are
investigated. Work in this direction has been done in travelling-wave
convection in bynary mixtures. In this case, the presence of distant
sidewalls in systems that are translation-invariant break the $O(2)$ symmetry
and the simplest possible symmetry-breaking effects are discussed 
when the system undergoes a Hopf bifurcation \cite{noblo1}. \par

	 This paper is organized as follows. In section 2 we discuss the 
properties of the physical system and we present the 
 equations describing the dynamics of the active primary modes  in 
this system. A linear study of these modes is performed in order to 
identify the possible secondary solutions that might bifurcate from them. 
In section 3 we analyse the conditions under which a nonlinear interaction 
between the secondary  solutions is possible, and a normal form reduction
of the equations is carried out.
After computing the coefficients of this normal form in terms of 
  parameter values for a $CO_2$ laser, we determine the two
possible scenarios . We discuss the kind of patterns
that could be observed under these conditions in section 4.
Section 5 sets out some conclusions and provides some guidelines for
  further experimental observation.

\section{The Model and Primary Bifurcations}

	The physical system is a $CO_2$ laser in a Fabry-Perot
 cavity. The active medium is contained in a cylindrical tube,
with a perfectly reflecting plane mirror at one end and a curved mirror
 with  partial
reflectivity at the other end. Physically the effective curvature of this mirror
 can be modified by inserting a passive optical device. Moreover, we are 
interested in the interaction among modes with nonzero angular momentum.
 This can only be achieved experimentally by placing an intracavity iris
 which inhibits the Gaussian mode \cite{green}.\par
 We therefore consider that the electric field can be expressed in terms
of the modes arising from a Hopf bifurcation from the zero solution:
  			
\begin {equation}
  E = P_1(r)L_1(z)e^{i \omega t}(z_1 e^{i l\theta}+z_2 e^{-i l\theta}), 
\end{equation}

where $P_1(r)$ and $L_1(z)$ account for the radial and longitudinal dependence
of the bifurcating modes, respectively, and $\theta$ is the angular 
variable. The integer $l$ denotes the angular 
momentum and  $\omega $ is  the temporal frequency of the modes.
The complex amplitudes $z_1$ and $z_2$ are governed by  the equations

\begin{eqnarray}
z_1' & = & \lambda  z_1 -A(z_1z_1^*+2z_2z_2^*)z_1+
           \epsilon z_2   \\
z_2' & = & \lambda  z_2 -A(2z_1z_1^*+z_2z_2^*)z_2+
           \epsilon z_1 
\end{eqnarray}

which are obtained by substituting the electric field $E$ into 
 the Maxwell-Bloch equations and truncating to third order \cite{solari}.
 The complex coefficients 
$\lambda$ and $A$ can be expressed in terms of the atomic inversion and 
decay rates, the detuning and a convolution between the pumping profile and
the spatial part of the modes. (The explicit form of these coefficients can be
in the Appendix). Note that the complex parameter $\epsilon$ 
(which carries information 
about the asymmetries in the laser parameters, such as anisotropies in 
the pumping or in the Brewster windows \cite{lopez}) accounts for the 
breaking of the $O(2)$ symmetry (see Appendix).  As the equations remain 
invariant under  the operation $z_1 \leftrightarrow z_2$, they preserve 
a $Z_2$ symmetry.\par

We substitute  $z_i= \rho_i e^{i \phi_i}$, $\epsilon = \rho_{\epsilon} 
e^{i \phi_{\epsilon}}$ and  $\lambda = \mu + i\omega$ in equations (2-3).
 After scaling by $A^r=Re(A)$ we arrive to the following system:

\begin{eqnarray}
\rho_1' & =& \mu \rho_1 - (\rho_1^2+2\rho_2^2)\rho_1+\rho_{\epsilon}
 \rho_2 \cos(\delta+
\phi_{\epsilon})\\
\rho_2' & =& \mu \rho_2 - (2 \rho_1^2+\rho_2^2)\rho_2+\rho_{\epsilon}
 \rho_1 \cos(\delta-
\phi_{\epsilon})\\
\delta' & =& \alpha (\rho_1^2-\rho_2^2)-\rho_{\epsilon}(\rho_1/\rho_2 
\sin(\delta-\phi_{\epsilon})
+\rho_2/\rho_1 \sin(\delta+\phi_{\epsilon}))
\end{eqnarray}
and a fourth equation for the evolution of the phase $\phi_1$ which is
uncoupled from these three equations. The variable 
$\delta=\phi_2-\phi_1$ is
the phase difference between the two modes. Notice that its dynamics is 
nontrivial because the broken symmetry term couples this phase difference with
the mode amplitudes. (For simplicity we use the same notation for
 the rescaled parameters). The information about the curvature, detuning, etc.
 contained in $A^r$ remains in the coefficient $\alpha=A^i/A^r$. \par
 
The primary bifurcations from zero in equations (4-6)
are four stationary solutions: two standing waves 
$SW_0 (\rho_1=\rho_2,\delta=0)$, $SW_{\pi}(\rho_1=\rho_2, \delta=\pi)$ 
and two solutions which are a mixture 
between traveling waves and standing waves
 $TW_1(\rho_1>\rho_2, \delta>0)$, $TW_2
 (\rho_1<\rho_2, \delta<0)$. These solutions give four stationary
patterns their stability depending on the parameter settings. By
an appropriate selection of the parameters in this situation a  good qualitative
 agreement with a previous experiment has been obtained \cite{dangelo}.\par

\section{Takens-Bogdanov Bifurcation}

	In this paper the attention is focussed in investigating the secondary 
bifurcations from the standing wave and on the new dynamics arising
from the nonlinear interaction between those secondary solutions. 
Before proceeding with the stability analysis
it is convenient to perform the following change of variables:

\begin{eqnarray}
\rho_1 & = & B \cos(\phi/2) \\
\rho_2 & = & B \sin(\phi/2). \nonumber
\end{eqnarray}

because  the $B$ direction is decoupled from the $(\phi, \delta)$
 plane in this representation (Fig.1). The standing wave solutions are simply 
$SW_{0,\pi}$:

\begin{displaymath}
B^2_{0,\pi}=2\frac{\mu\pm\rho_{\epsilon}\cos\phi_{\epsilon}}{3},
\; \phi_{0,\pi}=\pi/2,\;\delta_{0,\pi}=0,\pi
\end{displaymath}

 A linear stability analysis around these solutions leads to the following
 eigenvalues. In the $B$ direction the eigenvalue is simply
 $-2(\mu\pm \rho_{\epsilon}\cos(\phi_{\epsilon}))$, where the $\pm$ correspond
to the $SW_{0,\pi}$ solutions respectively. The eigenvalues corresponding
to the $(\phi,\delta)$ plane are obtained from the following matrix:

\begin{eqnarray}
 \left( \begin{array}{cc}
\frac{2}{3}\mu\mp \frac{4}{3}\rho_{\epsilon}\cos\phi_{\epsilon} & 
\pm 2 \rho_{\epsilon}\sin\phi_{\epsilon}  \\
\frac{2\alpha}{3}(\mu\pm\rho_{\epsilon}\cos\phi_{\epsilon})
\mp 2\rho_{\epsilon}\sin\phi_{\epsilon} & \mp 2\rho_{\epsilon}\cos\phi_{\epsilon}
\end{array}\right)  
\end{eqnarray}  
Now we determine the conditions for the appearance of a codimension-two (CT)
 point. 

\subsection{Codimension-Two Point}

The codimension of a bifurcation is the smallest dimension of a parameter 
space wich contains the bifurcation in a persistent way.
In our case the CT point is determined by a degenerated double zero
 eigenvalue in the matrix (8). This results from the interaction 
between a Pitchfork bifurcation and  a Hopf bifurcation. 
(The Pitchfork bifurcation gives rise to two new solutions 
($TW_{1,2}'$). The solution resulting from the Hopf bifurcation
is the so-called modulated wave ($MW$)). This CT point  
 is obtained when the parameters verify: 

\begin{eqnarray}
 \mu & = & \pm 5\rho_{\epsilon}\cos\phi_{\epsilon} \\
 \alpha^{-1} & = & - \tan(2\phi_{\epsilon}) 
\end{eqnarray} 

For a given $\alpha$, and as $\mu>0$ (the two modes were born),
these relations (9-10) lead to four solutions as sketched in Fig.2.
Solutions I and IV are associated with the $SW_0$ (sign + in eq. 9):
($I\equiv \phi_{\epsilon}^0; IV\equiv\phi_{\epsilon}^0 +3 \pi/2$),
while II and III ($II\equiv \phi_{\epsilon}^0 + \pi/2;
 III\equiv\phi_{\epsilon}^0 + \pi$) correspond to $SW_{\pi}$ (sign - in eq. 9).
 As eqs.(4-6) are invariant under the transformation
 $( \phi_{\epsilon};\rho_1(t),\rho_2(t),\delta(t) )\rightarrow 
(\phi_{\epsilon}+\pi;\rho_1(t),\rho_2(t),\delta(t)+\pi)$ 
($Z_2$-residual symmetry) the local dynamics around I and III is equivalent.
The same argument can be applied to solutions II and IV. \par

For the sake of simplicity  we analyze  only the local 
behavior around instability I(and IV) in the next subsection.
 (Calculations for cases II (and III) follow the same procedure).

\subsection{Normal Form Reduction}

Let us recall some general features of a dynamical system linearized around 
a stationary solution.
When some eigenvalues have zero real part, the flow near this fixed
point can be quite complicated. The linear space
spanded by the states corresponding to this null real part eigenvalues is 
known as {\it center eigenspace}.
The invariant manifold tangent to the center eigenspace constitutes the
{\it center manifold}. The local dynamics 'transverse' to this manifold
is relatively simple, since it is controlled by the 'fast' variables of the 
flow. The asymptotic behavior of the flow develops on the center manifold.
 The family of these behaviors that arise in the vicinity of
 this fixed point when the parameters are slightly varied is called
the {\it unfolding} of the bifurcation. The 'simplest' set of equations that 
reproduces generically  all these behaviors is called the 
{\it normal form} of the bifurcation. \par

 The unfolding of the bifurcation for solution I (dynamics near $SW_0$)
is described by the normal form:
\begin{eqnarray}
x' & = & y \\
y' & = & a x + b y \mp  x^3 - x^2y
\end{eqnarray}
that was first studied by Takens and Bogdanov \cite{guken,nobloc}.
it can show in fact two different behaviors depending the choice of
 the sign $\mp$.\par

Now we show that eqs.(4-6) reduce to eqs.(11-12) around $SW_0$
 if the conditions (9-10) for solution I  hold. \par
Let us note that in this case $\phi_{\epsilon}^0\in(0,\pi/2)$ 
for any given $\alpha$ (eqs.9-10).
We perform a last change of variables: $(B,\phi,\delta)\rightarrow(B,u,\delta)$
with $u=\frac{\rho_2^2-\rho_1^2}{B^2}=-\cos\phi$. The equations can be rewritten:
\begin{eqnarray}
B' & = & \mu B-\frac{1}{2}(3-u^2)B^3 +
\rho_{\epsilon}B(1-u^2)^{1/2}\cos\phi_{\epsilon}\cos\delta \\ 
u' & = & u(1-u^2)B^2 + 2\rho_{\epsilon}(1-u^2)^{1/2}
(\sin\phi_{\epsilon}\sin\delta -u\cos\phi_{\epsilon}\cos\delta ) \\
\delta ' & = & \alpha B^2 u - 
\frac{2\rho_{\epsilon}}{\sqrt{1-u^2}}(\cos\phi_{\epsilon}\sin\delta +
 u \sin\phi_{\epsilon }\cos{\delta })
\end{eqnarray}
\par
To obtain the unfolding of the bifurcation we consider
 for small variations of the parameters:
\begin{eqnarray}
\phi_{\epsilon} & = & \phi_{\epsilon}^0 + \beta \nonumber \\
\frac{\mu}{\rho_{\epsilon}} & = & 5\cos\phi_{\epsilon}^0+ q \nonumber
\end{eqnarray}
 The fast variable B is 'enslaved' by the other two variables and can
be adiabatically eliminated after assuming $B'=0$.
 Then the central manifold is two-dimensional and can be expressed 
as a function of (u,$\delta$). As usual in a local stability analysis, we will
keepup to third order terms. To obtain the normal form (eq.
(11-12)) the following change of variables 
$(u,\delta)\rightarrow(x=\frac{1}{2}
(\frac{u}{tag(\phi_{\epsilon}^0)}-
\delta),y=\frac{1}{2}(\frac{u}{tag(\phi_{\epsilon}^0)}+\delta))$
is introduced.
This allows to obtain the right linear part of the normal form.
We can rearrange the equations by means of another near-identity change 
on the new variables $(x, y)$. After these changes the normal form of the 
bifurcation reads as:
\par
\begin{eqnarray}
x' & = & y \\
y' & = & \mu_1 x + \mu_2 y + C x^3 + D x^2y
\end{eqnarray}
where
\begin{eqnarray}
\mu_1 & = & \frac{4}{3}(\frac{1+\alpha tag(\phi_{\epsilon}^0)}
{\cos(\phi_{\epsilon}^0)}q -
\frac{\alpha tag^3(\phi_{\epsilon}^0)+8\alpha tag(\phi_{\epsilon}^0)+7}
{tag(\phi_{\epsilon}^0)}\beta)\\
\mu_2 & = & \frac{2}{3}(\frac{1}{\cos(\phi_{\epsilon}^0)}q+
5tag^2(\phi_{\epsilon}^0)\beta)\\
C & = & \frac{2}{3}(2-\alpha tag(\phi_{\epsilon}^0)-
3\alpha tag^3(\phi_{\epsilon}^0))\\
D & = & \frac{-1}{6}(80\alpha tag(\phi_{\epsilon}^0)+32)\\
\alpha & = & \frac{tag^2(\phi_{\epsilon}^0)-1}{2tag(\phi_{\epsilon}^0)}
\end{eqnarray}

Notice that the coefficients in this normal form mainly depend
on the value of $\alpha$. As D is always positive the unfolding of the 
bifurcation depends mainly on C wich is a function mainly of $\alpha$.
 The unfolding is qualitatively different 
when C changes from negative to positive, i.e., for the conditions:

\begin{equation}
C=0\Rightarrow tag(\phi_{\epsilon}^*)=\sqrt{(5/3)}\Rightarrow \alpha_*=
-cotag(2\phi_{\epsilon}^*)=0.258
\end{equation}
\par

(The unfoldings corresponding to $C > 0$ and $C < 0$ will we described 
in detail in the next section). In the appendix
the dependence of $\alpha$ on the laser properties is 
calculated for different transverse modes. This calculations show that the 
transition point (eq. (23)) can be reached in a $CO_2$ laser
(Fig. 3).

\section{Pattern Dynamics in the Neighborhood of the Bifurcation}

In section III we obtained the conditions for the two posible kinds of 
T-B bifurcation that can
take place in the transverse  section of a $CO_2$ laser. Depending
on the sign of parameter C two qualitatively different scenarios are possible.
Now we describe the two different dynamics in some detail: 

\par
{\bf $ a)\;\;\; \alpha \in (-\infty,0.258)\equiv\phi_{\epsilon}\in 
(0,0.912)$}:\par
This corresponds to the case $C>0$ $(D<0)$. The unfolding near the bifurcation 
is presented in Figure 4a. From this unfolding the following general features
are read.
First, the solutios $SW_0$ (Fig. 5) as well as the two $TW'$
(created after a Pichtfork bifurcation from the $SW_0$) are unstable for
nearly all the parameter space.
The system almost ever evolves to some of the travelling solutions,
$TW_1\equiv(\rho_1,\rho_2,\delta)$ or $TW_2\equiv(\rho_2,\rho_1,-\delta)$.
So, one of the conjugate patterns in figure 5 
would be observed when we tune the parameter values around the bifurcation
point.
Second, the analysis of the full equations (4-6) shows that far away from the 
bifurcation point a limit cycle appears as a global
bifurcation from a heteroclinic connection between these patterns ($TW_{1,2}$).
This solution will look like a periodic alternancy between the
 two $TW_{1,2}$ due to the critical slowing down that takes place in the
 neighborhood of these points (Fig. 5). 
\par  
\par
{\bf $ b) \;\;\;\alpha\in (0.258,\infty )\equiv\phi_{\epsilon}\in (0.912,
\frac{\pi}{2})$}:\par
In this case $C<0$ $(D<0)$ and there is a strong change in the pattern
evolution (Fig. 4b). Near the bifurcation point the $SW_0$, weak 
oscillations $TW_{1,2}'\leftrightarrow SW_0$ (quasi-stationary patterns),
or a periodic alternancy
 $TW_1'\leftrightarrow SW_0\leftrightarrow TW_2'$ 
 will be observed. The latter corresponds to a limit cycle that grows in a
 global bifurcation from a homoclinic connection of the $SW_0$.
When we go far away in the parameter space a periodic alternancy between the
two $TW_{1,2}$ is found again. \par

Notice that the frequency associated to the oscillation between the patterns
 (Hopf bifurcation from $SW_0$) is of order 
$\omega_{osc}^2\sim\mid\mu_1\mid\sim\rho_{\epsilon}\mid q\mid$. So it is 
much slower than the temporal scale associated to the modes.
\par

\section{Conclusions}

In this article, a model for the evolution of transverse modes ($l=\pm 1$)
 in  a $CO_2$ laser 
has been studied. Due to the unavoidable anisotropies in the laser setup, some 
terms that break the natural $O(2)$ symmetry must be included.  
This model is quite succesful in predicting the stability of the primary 
solutions $SW$ found in recent experiments. The possible secondary solutions
have been determined.\par

The symmetry breaking term allows the possibility of a codimension-two
point bifurcating from the $SW$. This point appears when 
secondary solutions from a Pitchfork and a Hopf bifurcation 
interact simultaneously at the same point (Takens-Bogdanov bifurcation).  
A normal form reduction
procedure allowed to capture the main dynamical properties of the system
near that bifurcation. The calculations of the normal form coefficients for 
different realistic values in a $CO_2$ laser showed that 
two possible dynamical scenarios are possible. 

The intensity patterns that correspond to the dynamical interaction
between these modes ($l=\pm 1$) have been described for these two scenarios.
 Near the bifurcation one can get:
 stationary patterns (standing waves $SW$ or travelling waves $TW$)
 or oscillations between the $SW$ and the new conjugate patterns $TW'$.
Far away from the point bifurcation a global bifurcation of
a heteroclinic connection leads to a periodic alternancy between $TW$ 
patterns. These patterns look quasi-stationary due to a critical
slowing down near the $TW$ solutions.\par
Remarkably this "rich" dynamics cannot be observed without a 
symmetry-breaking term. It is important to stress that the interpretation 
of complicated pattern dynamics beyond the particular
problem studied here could lie on simple symmetry considerations. 
\par

\section{Acknowledgements}

We must acknowledge valuable comments and discussions with H. Solari (Buenos
Aires) and R. Gilmore (Philadelphia). This work has been partially supported by 
the DGICYT (Spanish Government) under grant PB90-0362.
One of us(R.L-R) acknowledges the Gobierno Foral de Navarra (Spain) for a 
research grant.

\section{Appendix}

It is clear that the analysis presented in this work strongly relies 
on the terms multipliying $\epsilon$ in Eqs. (2-3). An asymmetry in 
the pumping profile gives rise to this linear coupling between the equations
for the mode amplitudes $z_1$ and $z_2$. The dependence of $\epsilon$
in terms of laser parameters was derived in reference \cite{lopez}. The
result was:
\begin{equation}
\epsilon = \frac{K_{\epsilon}}{\beta - i\Omega_1} 
\end{equation}
where $K_\epsilon$ is the parameter governing the asymmetry in the pumping
profile ($K_{pumping}=K(r)+2K_{\epsilon}\cos(2\theta)$), $\beta$ is the rate
of decay of the atomic polarization and $\Omega_1$ is the slow temporal
frequency of the empty cavity mode. \par
In the following we compute the value of $\alpha=\frac{A^i}{A^r}$ 
for the primary bifurcating modes with angular momentum $l=\pm 1$. It is shown 
that the transition point $\alpha=0.258$ between the two possible T-B
bifurcations  can be obtained in a $CO_2$ laser. \par
The spatial coordinates of the problem are $(r,\theta,z)$ where $(r,\theta)$
corresponds to the transversal section of the cavity and $z$ to the longitudinal
direction $(z\in [-L,0])$. The new coordinates $(\xi,\theta,z)$ are introduced
\cite {solari}:

\begin{eqnarray}
\xi^2 & = & (\frac{r}{L})^2 ks(s^2+z^2)^{-1} \\
s & = & (\frac{2R_m}{L}-1)^{1/2}
\end{eqnarray}

where $s$ is the effective curvature, $L(\sim 1m)$ the cavity length, 
$R_m$ the curvature radius of the spherical mirror and
 $k=\frac{2\pi L}{\lambda}$ is the wave number. 
\par
The field is expanded (following \cite{solari}) 
in terms of empty cavity modes:
\begin{equation}
\left[ \begin{array}{c}
\Phi^+ \\
\Phi^-
\end{array} \right]  = \sum z_{\mu}
\left[ \begin{array}{c}
a_{\mu}^+ \\
a_{\mu}^-
\end{array} \right].  
\end{equation} 
This expansion is introduced in the Maxwell-Bloch equations 
(a brief review is done in \cite{lopez}), 
and projected on each mode. The temporal evolution for the 
cavity mode amplitudes $z_\alpha$ is then: 
\begin{equation}
z_{\alpha}'=\lambda_{\alpha}z_{\alpha}+ M_{\alpha \mu \nu \beta}z_{\mu}z_{\nu}^*
z_{\beta}+ h.o.t.
\end{equation}
where the expressions of the coefficients $\lambda_{\alpha}$, 
$M_{\alpha \mu \nu \beta}$ are given in \cite{solari}.
\par
The mode $a_\mu=(a_\mu^+,a_\mu^-)$ has the functional form:

\begin{equation}
a_{\mu}=(R_{\mu}(\xi)e^{-i\phi_\mu}e^{il_{\mu}\theta},
         R_{\mu}(\xi)e^{i\phi_\mu}e^{-il_{\mu}\theta})
\end{equation}

The coefficient $\phi_{mu}$ is given by:

\begin{eqnarray}
\phi_{\mu} & = & \Omega_{\mu}z-p_{\mu}\tan(\frac{z}{s}) 
\end{eqnarray}
with
\begin{eqnarray}
\Omega_{\mu} & = & n_{\mu}\pi + p_{\mu}\arctan(\frac{1}{s})-\delta
\end{eqnarray}

where ($p_{\mu},l_{\mu},n_{\mu}$) are three integers that characterize
 the mode: 
$p_{\mu}=2n_r+l_{\mu}+1$ is essentially the total tranverse energy of the 
laser beam  
($n_r$ is the radial quantum number), $l_{\mu}$ is the angular momentum 
around the z axis and $n_{\mu}$ is associated to the longitudinal behaviour of 
the mode. $R_{\mu}(\xi)$ is the radial dependence and $\delta=k\bmod 2\pi$ is
the detuning of the cavity.
\par
Making the operations indicated in \cite{solari}, we obtain the general form:  
\begin{eqnarray}
\lambda_{\alpha} & = & 
\left(-\chi + \frac{\bar{K}\beta}{\beta^2 + \Omega_{\alpha}^2}\right)
- i\Omega_{\alpha}
\left( 1 - \frac{\bar{K}}{\beta^2 + \Omega_{\alpha}^2}\right) \\
M_{\alpha \mu \nu \beta} & = & F(\Omega_{\mu},\Omega_{\nu},\Omega_{\beta})
\frac{L^2}{ks}\int\int\int K(\xi)R_{\alpha}(\xi)R_{\nu}(\xi)
R_{\mu}(\xi)R_{\beta}(\xi)\times \\
 & & [4\cos(\phi_{\alpha}+\phi_{\nu}-\phi_{\mu}-\phi_{\beta})
+ 2\cos(\phi_{\alpha}-\phi_{\nu}-\phi_{\mu}+\phi_{\beta})]\xi d\xi dz d\theta
\nonumber 
\end{eqnarray}
where the pumping profile $K(\xi)=\bar{K}$ is considered constant along
the whole beam width and $\chi$ gives account for the cavity losses. \par

After applying these procedures to our two mode interaction model: 

\begin{eqnarray}
z_1' & = & \lambda  z_1 - A(z_1z_1^*+2z_2z_2^*)z_1 + \epsilon z_1\\
z_2' & = & \lambda  z_2 - A(2z_1z_1^*+z_2z_2^*)z_2 + \epsilon z_2
\end{eqnarray}

we can identify $A=-M_{1111}$ that, according to expression (31) gives:
\begin{equation}
A=-M_{1111}=cte.\frac{2\beta}{\beta^2+\Omega_1^2}\frac{1}{\beta-i\Omega_1}
\int 6R_1(\xi)^4\xi d\xi
\end{equation}
These calculations have been performed for the primary bifurcating
modes $a_{\mu}=(p,l,n)$:

\begin{eqnarray}
b_1\equiv (2,1,0) & \rightarrow & R_1^2(\xi) \sim \xi^2e^{-\xi^2} \\
c_1\equiv (4,1,0) & \rightarrow & R_1^2(\xi) \sim (2-\xi^2)^2\xi^2e^{-\xi^2} \\
b_1'\equiv (2,1,1) & \rightarrow & R_1^2(\xi) \sim \xi^2e^{-\xi^2}
\end{eqnarray}

In general the coefficient $\alpha=A^i/A^r$ depends on the cavity curvature s
and on the detuning $\delta$. We plot the dependency of $\alpha$ on
 s and $\delta$ for the primary modes in figure 3.  $\alpha$ decreases
when the curvature (or detuning) increases and the transition value
 $\alpha=0.258$ can be reached for a sufficiently high curvature s.

\newpage
\begin{center}{\bf Figure Captions}\end{center}\par
{\bf 1.} Sketch of the change of variables (7) performed in the
 equations (4-6). In these variables the dynamics of interest takes place
near the ($\delta$, $\phi$) plane (center manifold). \par

{\bf 2.} The four possible solutions of the equation system (9-10) 
(under the condition $\mu >0$) where
a Takens-Bogdanov bifurcation takes place. They are shown in the angular
$\phi_{\epsilon}$ space. The dynamics around all them is equivalent. \par

{\bf 3.} (a) Plot of the value of $\alpha$ as the curvature s of the
cavity is varied for three different modes: $b_1=(2,1,0)$, $c_1=(4,1,0)$ and
$b_1'=(2,1,1)$. The detuning $\delta$ is fixed: $\delta=0$. \newline
(b) Plot of $\alpha$ on s for the mode $b_1=(2,1,0)$ when the detuning
 takes the values: $\delta=1$, $\delta=2$ and $\delta=3$. \newline
(c) Plot of $\alpha$ on $\delta$ for three modes with different
longitudinal behavior: $b_1=(2,1,0)$, $b_1'=(2,1,1)$ and $b_1''=(2,1,2)$.
(The curvature is fixed: $s=20$).\newline
 The transition explained in relation (23)
is obtained for $\alpha=0.258$. (See appendix for the notation).\par 

{\bf 4.} Unfolding diagrams for the two kind of Takens-Bogdanov
 bifurcation (see the normal form 16-17)
that can be achieved: squares$\equiv$ $TW'_{1,2}$ (or $TW_{1,2}$),
circles$\equiv SW_0$. The outer figures of the diagram are obtained simulating
 the total number of equations (4-6) far from the instability.\newline
(a) For $\alpha<0.258$ where $C>0,D<0$, \newline
(b) For $\alpha>0.258$ where $C<0,D<0$. \par

{\bf 5.} Patterns that can be observed: $TW_1$ (or $TW_1'$ if it is close
to $SW_0$), $SW_0$ and  $TW_2$ (or $TW_2'$). The arrows represent
the periodic alternancy $TW_1'$ $\leftrightarrow$ 
$SW_0$ $\leftrightarrow$ $TW_2'$ that takes place for $C<0$.
It is also posible another periodic alternancy 
$TW_1'$ $\leftrightarrow$ $TW_2'$ or $TW_1$ $\leftrightarrow$ $TW_2$ 
for $C>0$. \par

\end{document}